\documentclass[superscriptaddress,aps,prl,showkeys,showpacs,twocolumn,10pt]{revtex4-1}

\usepackage{graphicx}
\usepackage{dcolumn}
\usepackage{bm}
\usepackage[OT1,T1]{fontenc}

\newcommand{\scaleFactor}{0.4}

\begin{document}
\title{Search for $\eta$-mesic $^4$He \\ with the WASA-at-COSY detector}
\date{\today}

\newcommand*{\IKPUU}{Division of Nuclear Physics, Department of Physics and 
 Astronomy, Uppsala University, Box 516, 75120 Uppsala, Sweden}
\newcommand*{\ASWarsN}{Department of Nuclear Physics, National Centre for 
 Nuclear Research, ul.\ Hoza~69, 00-681, Warsaw, Poland}
\newcommand*{\IPJ}{Institute of Physics, Jagiellonian University, ul.\ 
 Reymonta~4, 30-059 Krak\'{o}w, Poland}
\newcommand*{\PITue}{Physikalisches Institut, Eberhard--Karls--Universit\"at 
 T\"ubingen, Auf der Morgenstelle~14, 72076 T\"ubingen, Germany}
\newcommand*{\MS}{Institut f\"ur Kernphysik, Westf\"alische 
 Wilhelms--Universit\"at M\"unster, Wilhelm--Klemm--Str.~9, 48149 M\"unster, 
 Germany}
\newcommand*{\ASWarsH}{High Energy Physics Department, National Centre for 
 Nuclear Research, ul.\ Hoza~69, 00-681, Warsaw, Poland}
\newcommand*{\IITB}{Department of Physics, Indian Institute of Technology 
 Bombay, Powai, Mumbai--400076, Maharashtra, India}
\newcommand*{\IKPJ}{Institut f\"ur Kernphysik, Forschungszentrum J\"ulich, 
 52425 J\"ulich, Germany}
\newcommand*{\JCHP}{J\"ulich Center for Hadron Physics, Forschungszentrum 
 J\"ulich, 52425 J\"ulich, Germany}
\newcommand*{\Bochum}{Institut f\"ur Experimentalphysik I, Ruhr--Universit\"at 
 Bochum, Universit\"atsstr.~150, 44780 Bochum, Germany}
\newcommand*{\ZELJ}{Zentralinstitut f\"ur Elektronik, Forschungszentrum 
 J\"ulich, 52425 J\"ulich, Germany}
\newcommand*{\Erl}{Physikalisches Institut, 
 Friedrich--Alexander--Universit\"at Erlangen--N\"urnberg, 
 Erwin--Rommel-Str.~1, 91058 Erlangen, Germany}
\newcommand*{\ITEP}{Institute for Theoretical and Experimental Physics, State 
 Scientific Center of the Russian Federation, Bolshaya Cheremushkinskaya~25, 
 117218 Moscow, Russia}
\newcommand*{\Giess}{II.\ Physikalisches Institut, 
 Justus--Liebig--Universit\"at Gie{\ss}en, Heinrich--Buff--Ring~16, 35392 
 Giessen, Germany}
\newcommand*{\IITI}{Department of Physics, Indian Institute of Technology 
 Indore, Khandwa Road, Indore--452017, Madhya Pradesh, India}
\newcommand*{\HepGat}{High Energy Physics Division, Petersburg Nuclear Physics 
 Institute, Orlova Rosha~2, 188300 Gatchina, Russia}
\newcommand*{\HISKP}{Helmholtz--Institut f\"ur Strahlen-- und Kernphysik, 
 Rheinische Friedrich--Wilhelms--Universit\"at Bonn, Nu{\ss}allee~14--16, 
 53115 Bonn, Germany}
\newcommand*{\Katow}{August Che{\l}kowski Institute of Physics, University of 
 Silesia, Uniwersytecka~4, 40-007, Katowice, Poland}
\newcommand*{\IFJ}{The Henryk Niewodnicza{\'n}ski Institute of Nuclear Physics, 
 Polish Academy of Sciences, 152~Radzikowskiego St, 31-342 Krak\'{o}w, Poland}
\newcommand*{\NuJINR}{Dzhelepov Laboratory of Nuclear Problems, Joint 
 Institute for Nuclear Physics, Joliot--Curie~6, 141980 Dubna, Russia}
\newcommand*{\KEK}{High Energy Accelerator Research Organisation KEK, Tsukuba, 
 Ibaraki 305--0801, Japan}
\newcommand*{\IMPCAS}{Institute of Modern Physics, Chinese Academy of 
 Sciences, 509 Nanchang Rd., 730000 Lanzhou, China}
\newcommand*{\ASLodz}{Department of Cosmic Ray Physics, National Centre for 
 Nuclear Research, ul.\ Uniwersytecka~5, 90--950 {\L}\'{o}d\'{z}, Poland}

\author{P.~Adlarson}    \affiliation{\IKPUU}
\author{W.~Augustyniak} \affiliation{\ASWarsN}
\author{W.~Bardan}      \affiliation{\IPJ}
\author{M.~Bashkanov}   \affiliation{\PITue}
\author{T.~Bednarski}   \affiliation{\IPJ}
\author{F.S.~Bergmann}  \affiliation{\MS}
\author{M.~Ber{\l}owski}\affiliation{\ASWarsH}
\author{H.~Bhatt}       \affiliation{\IITB}
\author{M.~B\"uscher}   \affiliation{\IKPJ}\affiliation{\JCHP}
\author{H.~Cal\'{e}n}   \affiliation{\IKPUU}
% ...

\newcommand*{\SU}{Department of Physics, Stockholm University, 
 Roslagstullsbacken~21, AlbaNova, 10691 Stockholm, Sweden}
\newcommand*{\Mainz}{Institut f\"ur Kernphysik, Johannes 
 Gutenberg--Universit\"at Mainz, Johann--Joachim--Becher Weg~45, 55128 Mainz, 
 Germany}
\newcommand*{\UCLA}{Department of Physics and Astronomy, University of 
 California, Los Angeles, California--90045, U.S.A.}
\newcommand*{\Bern}{Albert Einstein Center for Fundamental Physics, 
 Fachbereich Physik und Astronomie, Universit\"at Bern, Sidlerstr.~5, 
 3012 Bern, Switzerland}
\newcommand*{\Sheff}{Department of Physics and Astronomy, University of 
 Sheffield, Hounsfield Road, Sheffield, S3 7RH, United Kingdom}

\collaboration{WASA-at-COSY Collaboration}\noaffiliation

\begin{abstract}
An exclusive measurement
of the excitation function for the $dd \rightarrow ^3$He$ p \pi^-$ reaction 
was performed at the Cooler Synchrotron COSY-J\"ulich with the WASA-at-COSY detection system.
The data were taken during a slow acceleration of the beam
from 2.185\,GeV/c to 2.400\,GeV/c crossing the kinematic threshold
for the $\eta$ meson production in the $dd \rightarrow ^4$He$\,\eta$ reaction at 2.336~GeV/c.
The corresponding excess energy with respect to the $^4$He$-\eta$ system varied from -51.4~MeV to 22~MeV.
The integrated luminosity in the experiment was determined using the $dd \rightarrow ^3$He$ n$ reaction.
The shape of the excitation function for the $dd \rightarrow ^3$He$ p \pi^{-}$ was examined.
No signal of the $^4$He$-\eta$ bound state was observed.
An upper limit for the cross-section for the bound state formation and decay in the process
$dd \rightarrow (^4$He$-\eta)_{bound} \rightarrow ^3$He$ p \pi^{-}$,
was determined on the 90 \% confidence level and it varies from 20~nb to 27~nb for the bound state width ranging from 5~MeV to 35~MeV, respectively.
\end{abstract}


\maketitle
\section{\label{sec:intro}Introduction}
Neutral mesons such as e.g. $\eta, K,\omega,\eta'$ can potentially form bound states
with atomic nuclei.
In this case the binding is exclusively due to the strong interaction and the bound state or {\em mesic nucleus}
- can be considered as a meson captured in the mean field of the nucleons.
Due to the strong attractive $\eta$-nucleon interaction~\cite{wycech,moskal}, the $\eta$-mesic nuclei are some of the most promising candidates for such states.

Experimental confirmation of the existence of $\eta$-mesic nuclei would be interesting on its own but it would be also
valuable for investigations of the $\eta-N$ interaction and for the study of in-medium properties
of the $N^{*}$ resonance~\cite{jido} and of the $\eta$ meson~\cite{osetNP710}.
It could also help to determine the flavour singlet
component of the $\eta$ wave function~\cite{basssymposium}.

The existence of $\eta$-mesic nuclei was postulated in 1986 by Haider and Liu \cite{liu2}.
Experimental searches have been performed by several past experiments~\cite{lampf,lpi,gsi,gem,mami} while ongoing investigations continue at COSY~\cite{moskalsymposium,jurek-he3,timo,meson08, jurekmeson08, Magda2}, JINR~\cite{jinr}, J-PARC~\cite{fujiokasymposium}, MAMI~\cite{mami2} and are planned at GSI~\cite{gsiplan}.
Many promising indications where reported, however, so far there is no direct experimental confirmation of the existence of mesic nuclei.

In the region of the light nuclei systems such as e.g. $\eta$-He the observation of a strong enhancement in the total production cross-section and the phase variation of the scattering amplitude in the close-to-threshold region provided strong evidence for the existence of a pole in the scattering matrix which can correspond to a bound state~\cite{wilkin2}.
In particular, a very strong final state interaction (FSI) is observed
in the $dd \rightarrow ^4$He$ \eta$ reaction close to kinematic threshold
and is interpreted as a possible indication of $^4$He$-\eta$ bound state~\cite{Willis97}. This
suggests, that the $^4$He$-\eta$ system is a good candidate for the experimental study of a possible binding. This conclusion is strengthened  by the predictions in Reference~\cite{wycech}.

However, as stated in Reference~\cite{liu3,haider2}, 
the theoretical predictions for the width and binding energy of the $\eta$-mesic nuclei 
are strongly dependent on the subthreshold $\eta$-nucleon interaction which is not well understood.
Therefore, direct measurements which could confirm the existence of the bound state are mandatory.

Taking into account the above arguments and the fact that in the light nuclei systems the bound states are expected to be much narrower compared to
the case of the heavy nuclei~\cite{Oset}, we performed a search for $\eta$-mesic $^{4}$He 
at the Cooler Synchrotron COSY-J\"ulich with the WASA-at-COSY detector~\cite{prop186}.

\section{Experiment}

The experiment was based on the measurement of the excitation function
of the $dd \rightarrow ^3$He$ p \pi^{-}$ reaction for energies in the vicinity of the $\eta$ production
threshold and on the selection of events with low ${^3}$He center-of-mass (c.m.) momenta.
In the case of existence of the $^4$He$-\eta$ bound state we expected to observe
a resonance-like structure in the excitation function below the threshold for the production of the $^4$He$-\eta$ system.
The structure would indicate a  $dd \rightarrow (^4$He$-\eta)_{bound} \rightarrow ^3$He$ p \pi^{-}$ 
reaction appearing on the continuous
background originating from the direct $dd \rightarrow ^3$He$ p \pi^{-}$ process.

We expect that the decay of such a state proceeds via 
the absorption of the $\eta$ meson on one of the 
nucleons in the $^4$He nucleus leading to the excitation of the  $N^{*}$ (1535)
resonance which subsequently decays in pion-nucleon pair.
The remaining three nucleons play the role of spectators and they are likely to bind forming  a $^3$He
or $^3$H nucleus.
In the case of a similar system, the  ${^{4}_{\Lambda}\mbox{He}}$ hypernucleus, it was observed that in the $\pi^{-}$ decay channel the decay mode $ {^{4}_{\Lambda}\mbox{He}} \rightarrow ^3$He$ p \pi^{-}$ is dominant~\cite{Fet}. 

According to the discussed model, there exist four equivalent decay channels
of the $(^4$He$-\eta)_{bound}$ state:
\begin{itemize}
\item   $(^4$He$-\eta)_{bound} \rightarrow ^3$He$ p \pi^{-}$
\item   $(^4$He$-\eta)_{bound} \rightarrow ^3$He$ n \pi^{0}$
\item   $(^4$He$-\eta)_{bound} \rightarrow {^3\mbox{H}} p \pi^{0}$
\item   $(^4$He$-\eta)_{bound} \rightarrow {^3\mbox{H}} n \pi^{+}$
\end{itemize}

In the reported experiment we concentrated on the $^3$He$ p \pi^{-}$ decay mode.

The WASA-at-COSY detector is described in detail in Ref.~\cite{wasa,WasaComm}. 
It consists of two main parts: the Forward Detector dedicated to the measurement
of forward-scattered projectiles and target-recoils, and the Central Detector,
optimized for measuring of photons, electrons and pions originating from decays of mesons and excited baryonic states.
The forward part consists of several layers of plastic scintillators allowing for
particle identification on the basis of the $\Delta$E-E and $\Delta$E-$\Delta$E information
and a proportional drift chamber providing track coordinates.
The Central Detector is composed of an electromagnetic calorimeter, 
a cylindrical drift chamber and a barrel of plastic scin\-ti\-lla\-tors. 
A superconducting solenoid provides a magnetic field for momentum determination of the tracks of charged particles measured in the drift chamber.
The scintillators provide fast signals for the first level trigger, and together with the drift chamber and the calorimeter, are used for charged particle identification via $\Delta$E-p and $\Delta$E-E methods. WASA-at-COSY uses
an internal target system which provides pellets of frozen hydrogen or deuterium.
During the present experiment the cooling system of the superconducting solenoid  was broken and, therefore, no magnetic field was provided. 

During the experimental run the momentum of the deuteron beam was varied continuously within each acceleration cycle
from  2.185~GeV/c to 2.400~GeV/c, crossing the kinematic threshold for $\eta$ production in the $dd \rightarrow ^4$He$\,\eta$ reaction at 2.336~GeV/c.
This range of beam momenta corresponds to a variation of $^4\mbox{He}-\eta$  excess energy  from -51.4~MeV to 22~MeV.

\begin{figure}
\begin{center}
      \scalebox{\scaleFactor}
         {
         \includegraphics{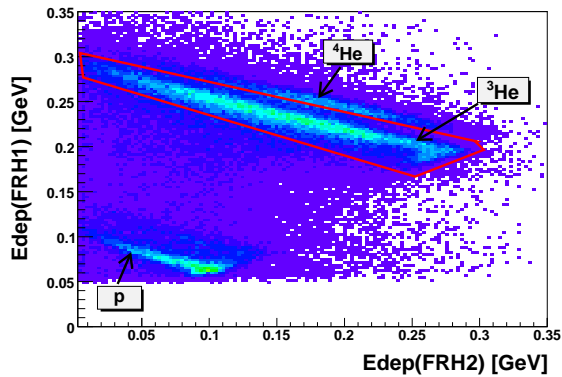}
  	 }

\caption{\label{he3selection_second_p}(Color online) Experimental spectrum  of the energy losses in first two layers of the Forward Range Hodoscope. 
The area used for $^{3}{\mbox{He}}$ identification is indicated by the red line.
The empty area below 0.05\,GeV in the Edep(FRH1) distribution is due to the preselection condition. The regions corresponding to protons, $^{3}{\mbox{He}}$ and $^{4}{\mbox{He}}$
are clearly visible.}
\end{center}
\end{figure}

The identification of the $^{3}{\mbox{He}}$ was conducted using the  $\Delta E- \Delta E$
technique, comparing the energy losses in two layers of the Forward Range Hodoscope (Fig.~\ref{he3selection_second_p}).

\begin{figure}
\begin{center}

      \scalebox{\scaleFactor}
         {
         \includegraphics{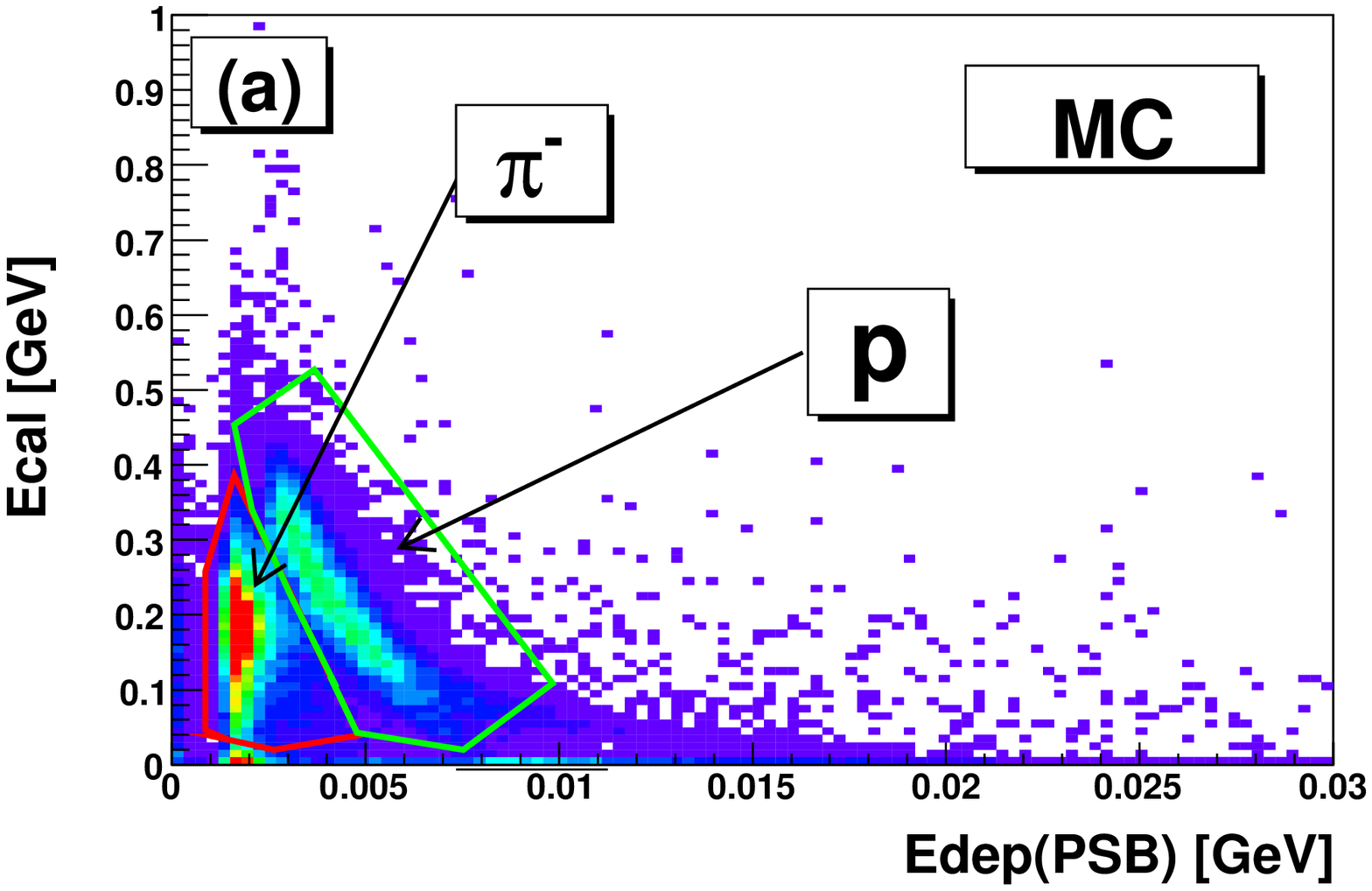}
  	 }
      \scalebox{\scaleFactor}
         {
         \includegraphics{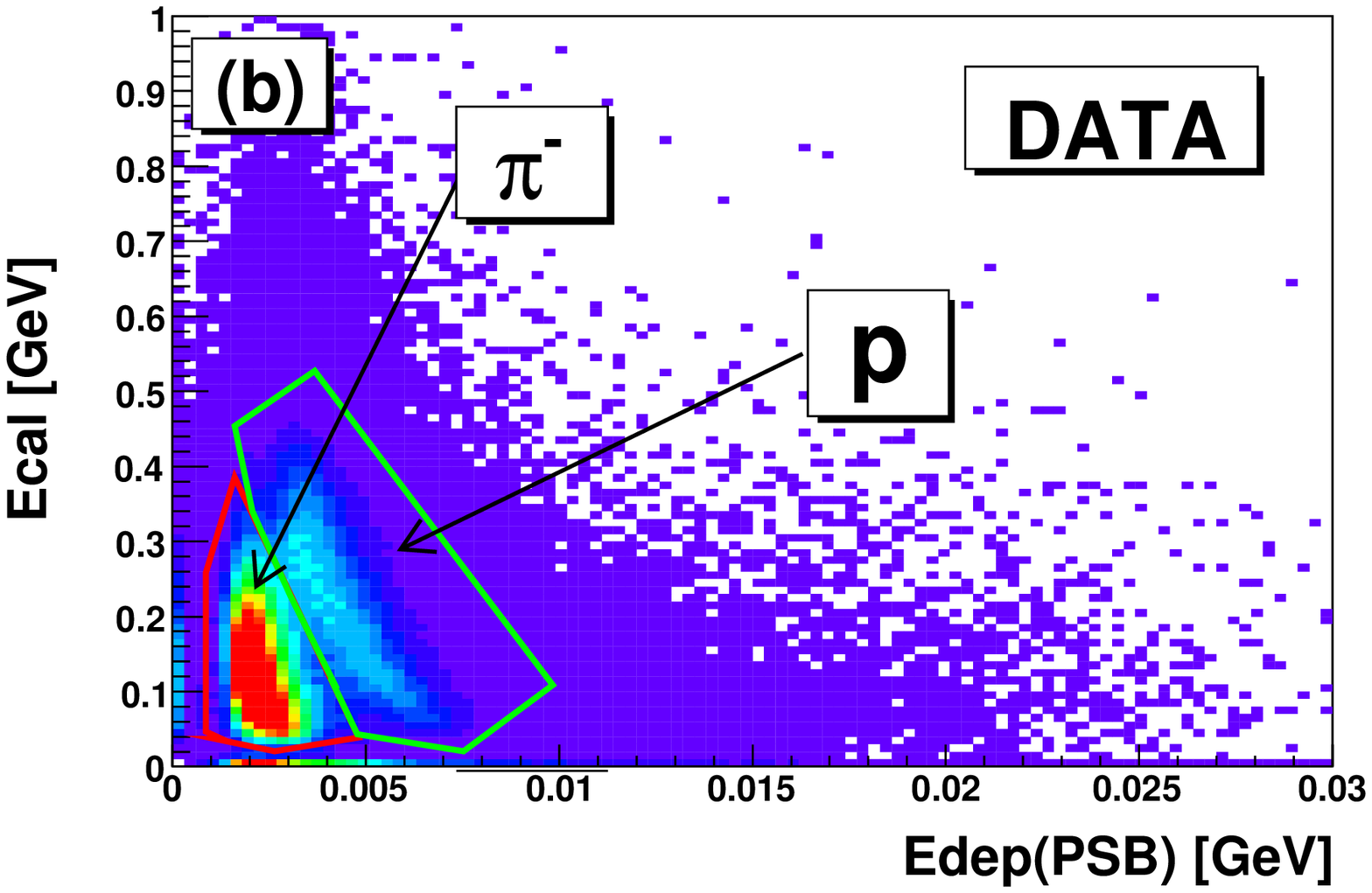}
  	 }
\caption[Identification of $p$ and $pi^{-}$.]{\label{ecal_vs_ps_p}(Color online) Comparison of the Monte Carlo simulation (a) and the experimental spectrum (b) of the energy loss in the Plastic Scintillator Barrel ($x$-axis) combined with the energy deposited in the Electromagnetic Calorimeter ($y$-axis). The green and red curves represent
the applied graphical condition to separate protons and pions.}
\end{center}
\end{figure}

The energy loss in the Plastic Scintillator Barrel was combined with the energy deposited
in the Electromagnetic Calorimeter to identify protons and pions (Fig.~\ref{ecal_vs_ps_p}).

\begin{figure}
\begin{center}
      \scalebox{\scaleFactor}
         {
              \includegraphics{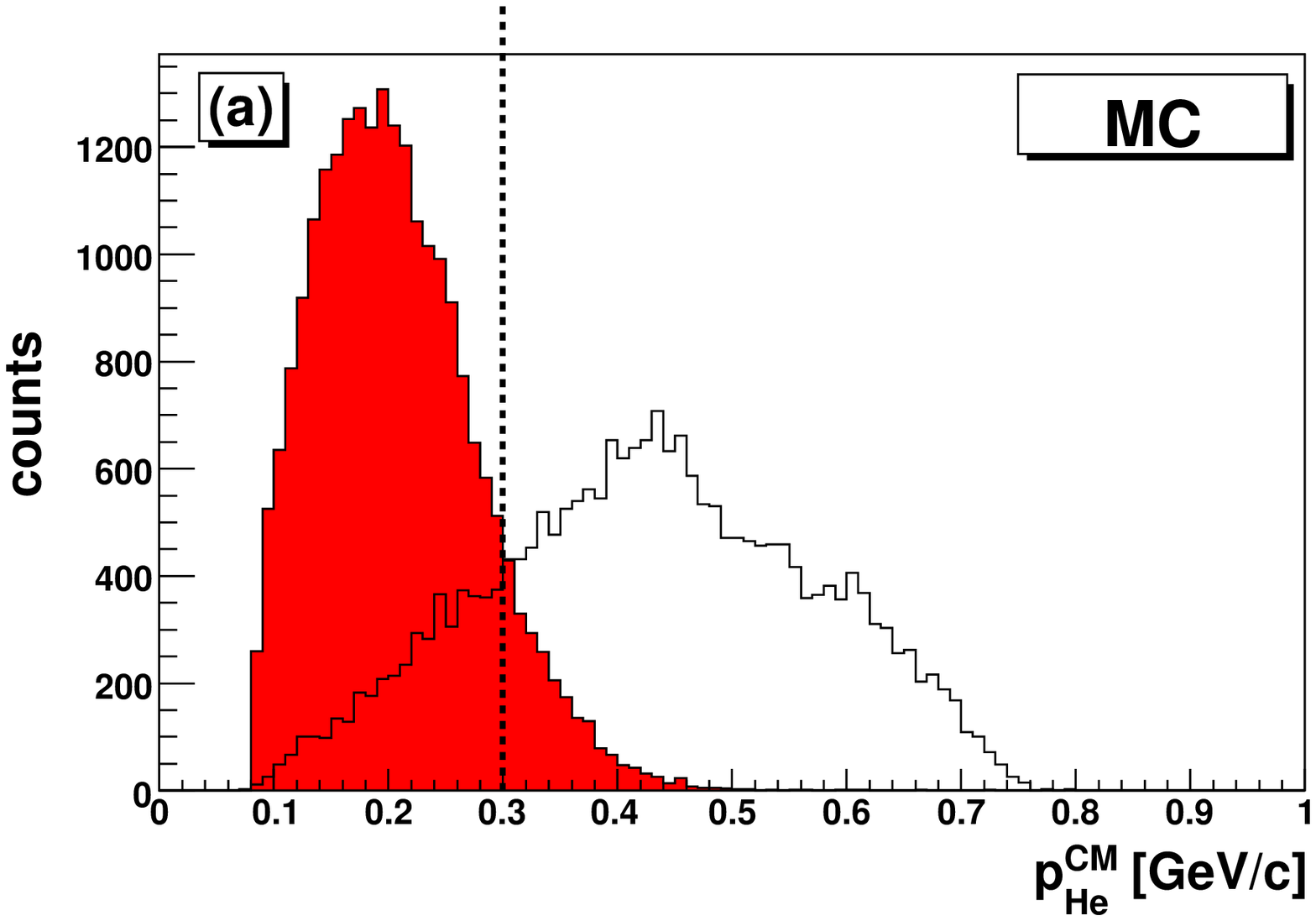}
         }
      \scalebox{\scaleFactor}
         {
              \includegraphics{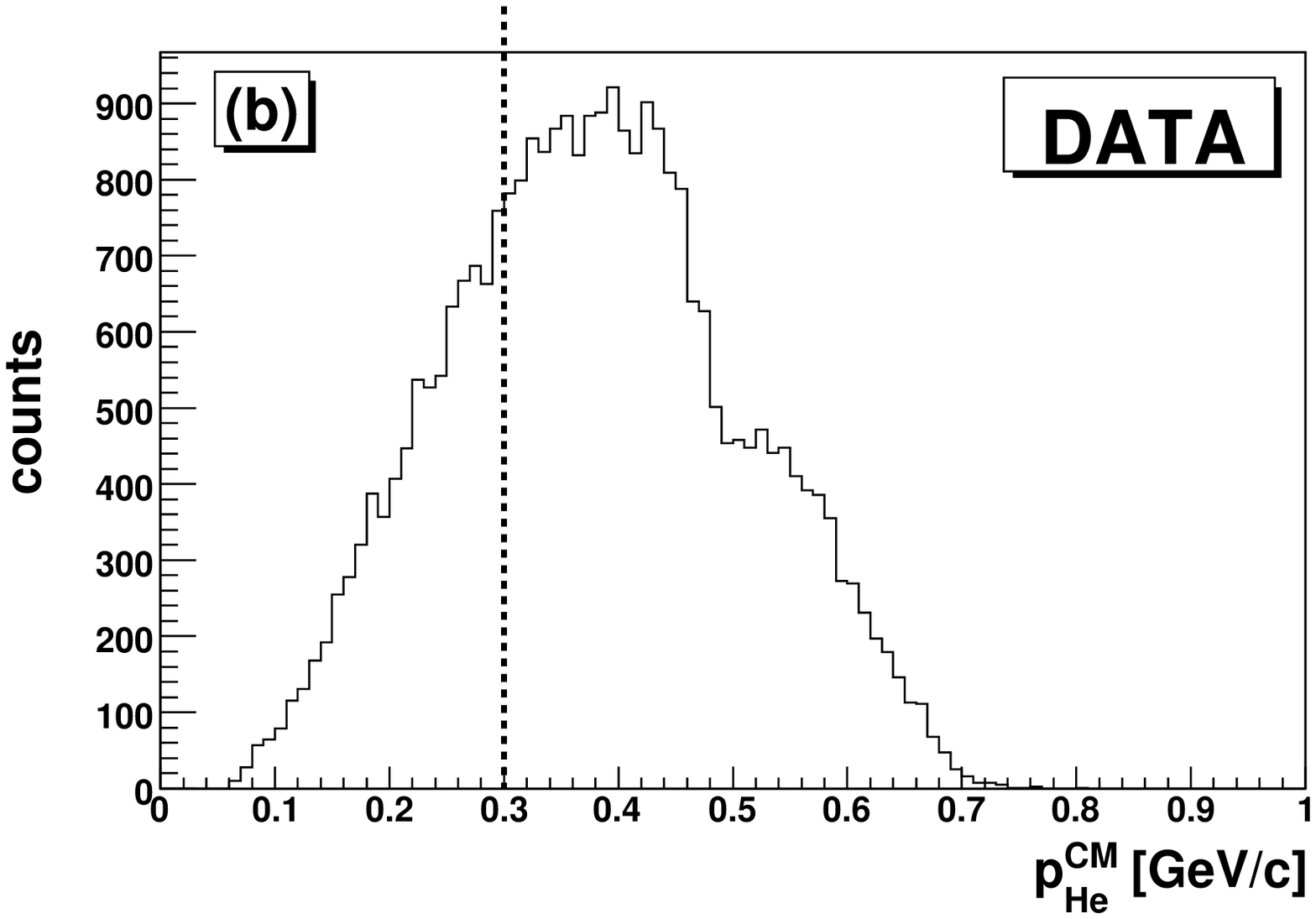}
         }

\caption[Distribution of the $^3$He momentum in c.m.]{\label{pmomc.m._p}(Color online) (Upper plot) Distribution of the $^3$He momentum in 
the c.m. system simulated for the processes leading to the creation
of the  $^4$He$\eta$ bound state:
$dd \rightarrow (^4{\mbox{He}} \eta)_{bound} \rightarrow ^3$He$p\pi^{-}$ (red area)
and of the phase-space $dd \rightarrow ^3$He$ p\pi^{-}$ reaction (black line).
The simulation was done for a momentum of the deuteron beam of 2.307~GeV/c.
The Fermi momentum parametrization was taken from \cite{VH}.
(plot b) Experimental distribution of the $^3$He momentum in the c.m. system.
In both plots the dashed line demarcates the "signal-poor" and the "signal-rich" regions.
The decrease of the counts at 0.48~GeV/c is due to the geometry of the border of the barrel and the end-caps of the Scintillator Barrel detector 
which was used in the $p-\pi^{-}$ identification process.
This region has no relevance  in the next steps of the analysis.
}
\end{center}
\end{figure}

The outgoing $^3\mbox{He}$ nucleus plays the role of a spectator and, therefore,
we expect that its momentum in the c.m. frame is relatively low and can be approximated 
by the Fermi momentum distribution of nucleons inside the $^4\mbox{He}$ nucleus. 
This signature  allows us to suppress background from reactions leading to the
$^3$He$ p \pi^{-}$ final state but proceeding without formation of the intermediate
$(^4$He$-\eta)_{bound}$ state and, therefore, resulting on the average in much higher
c.m. momenta of $^3\mbox{He}$ (see Fig.~\ref{pmomc.m._p}).

\begin{figure}
\begin{center}
  \scalebox{\scaleFactor}
  {
    \includegraphics{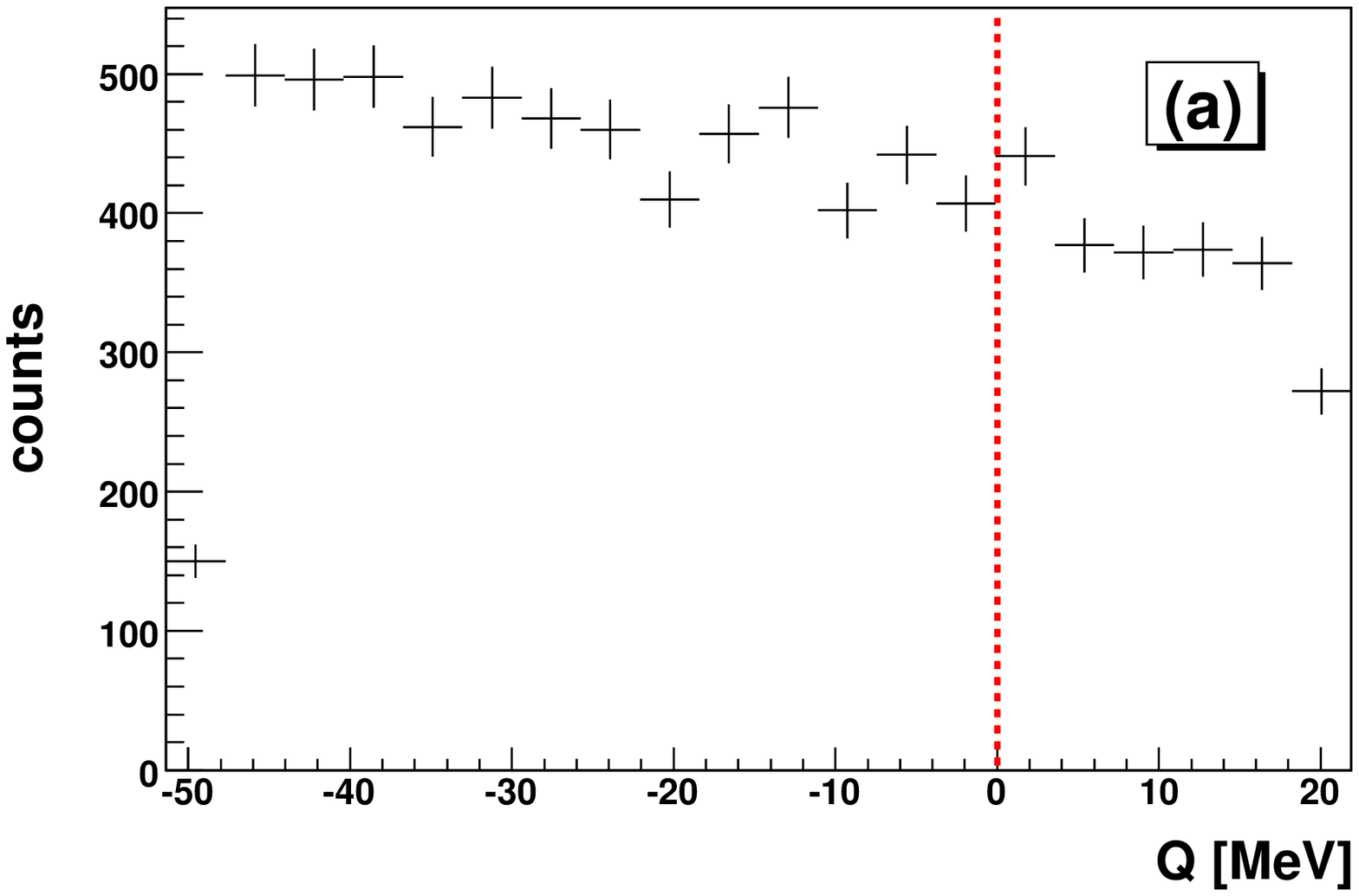}
  }
  \scalebox{\scaleFactor}
  {
    \includegraphics{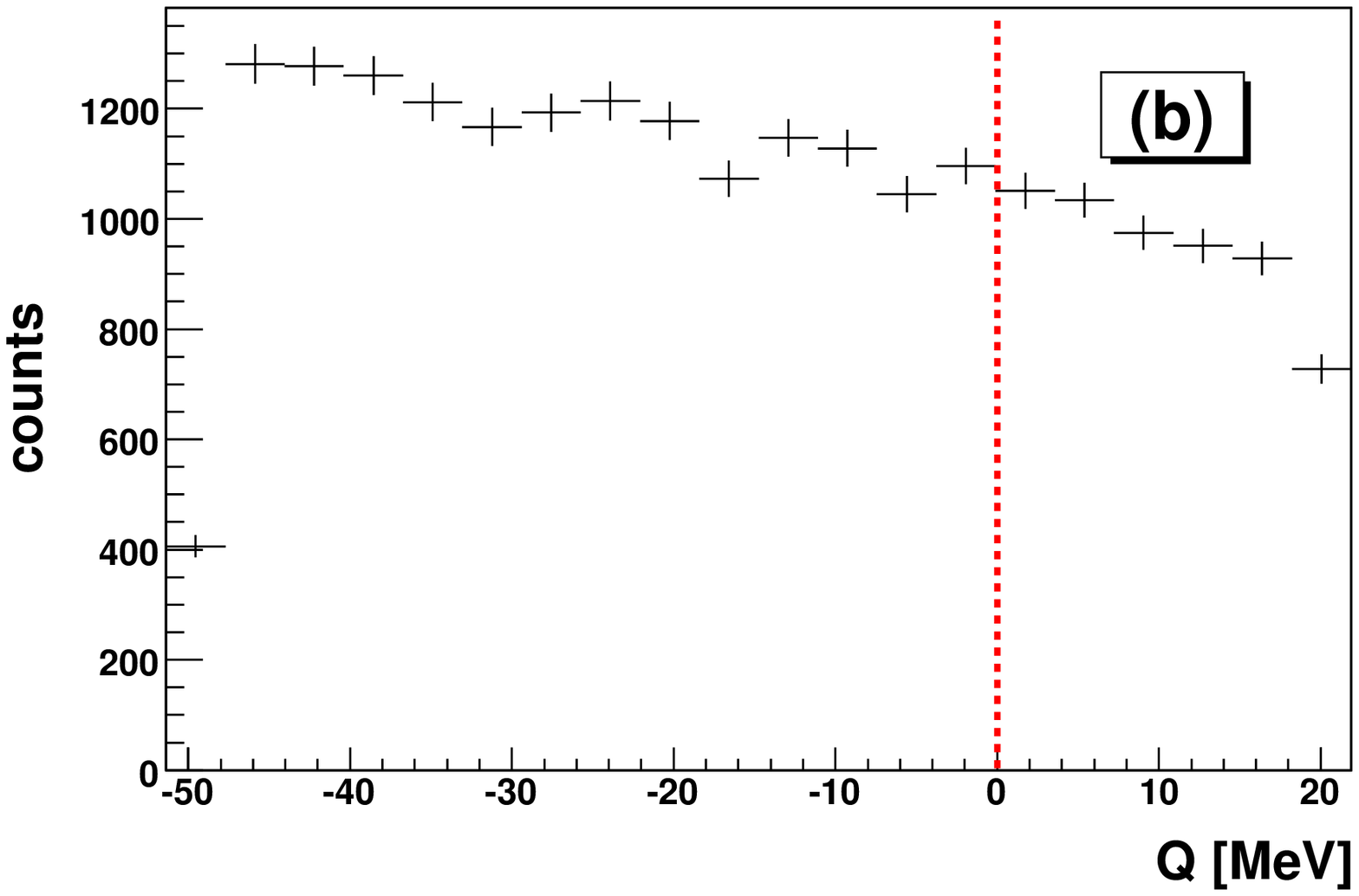}
  }
  \scalebox{\scaleFactor}
  {
    \includegraphics{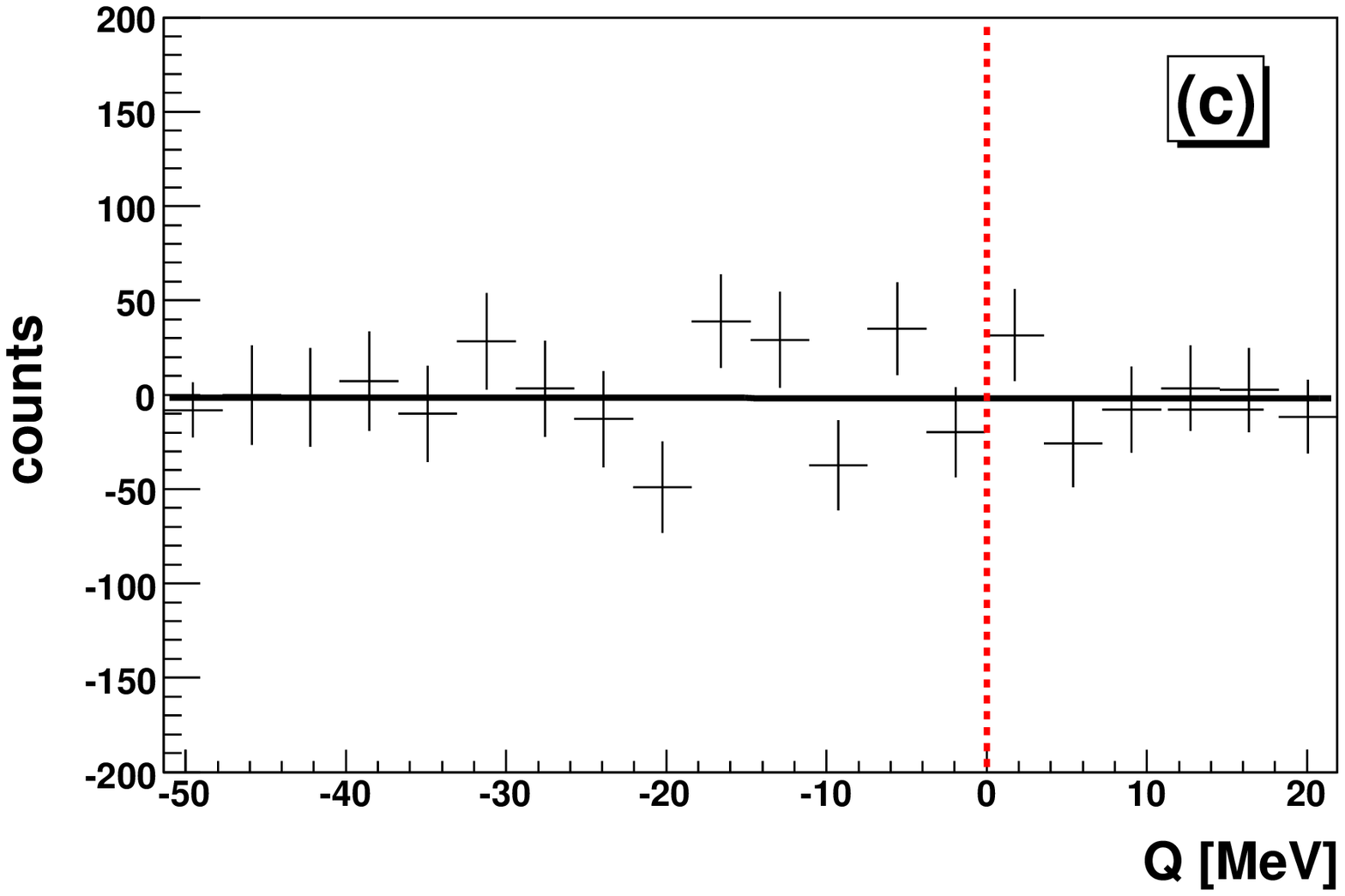}
  }
\caption[Not normalized excitation functions]{\label{hExcitFuncc.m._mom_exp_bad_p}(Color online) Excitation function for the $dd \rightarrow ^3$He$ p \pi^{-}$ reaction for the "signal-rich" region corresponding to $^3$He momentum below 0.3\,GeV/c (panel a) and the "signal-poor" region with  $^3$He momentum above 0.3\,GeV/c (panel b).
Difference of the excitation functions for the "signal-rich" and "signal-poor" regions after the normalization
to the second bin of Q is shown in the panel c. The black solid line represents a straight line fit.
The threshold of $^4$He$-\eta$  is marked by the vertical dashed line.}
\end{center}
\end{figure}

Therefore, we compare the excitation functions for the $dd \rightarrow ^3$He$ p \pi^{-}$ reaction
from the  "signal-rich" region
corresponding to the $^3$He c.m. momenta below 0.3\,GeV/c and the "signal-poor" region for the $^3$He c.m. momenta above 0.3\,GeV/c.
The number of events as a function of the beam momentum is shown in Fig.~\ref{hExcitFuncc.m._mom_exp_bad_p}. 
At this stage of the analysis the excitation function is not normalized to the luminosity and it is not corrected for reconstruction efficiency.
The obtained functions for both regions are smooth and no clear signal, which could be interpreted as a resonance-like
structure, is visible.
We checked also for possible structures in the difference between the excitation functions
for the "signal-rich" and "signal-poor" region.
We multiplied the function for the "signal-poor" region by a factor chosen in such a way,
that the difference of the two functions for the second bin of Q is equal to zero.
This difference  is presented in Fig.~\ref{hExcitFuncc.m._mom_exp_bad_p} (c) in order to examine 
the shape of the excitation function before any 
further selection criteria are applied. 
The obtained dependence is flat and is consistent with zero. No resonance structure is visible.

\begin{figure}
\begin{center}
      \scalebox{\scaleFactor}
         {
              \includegraphics{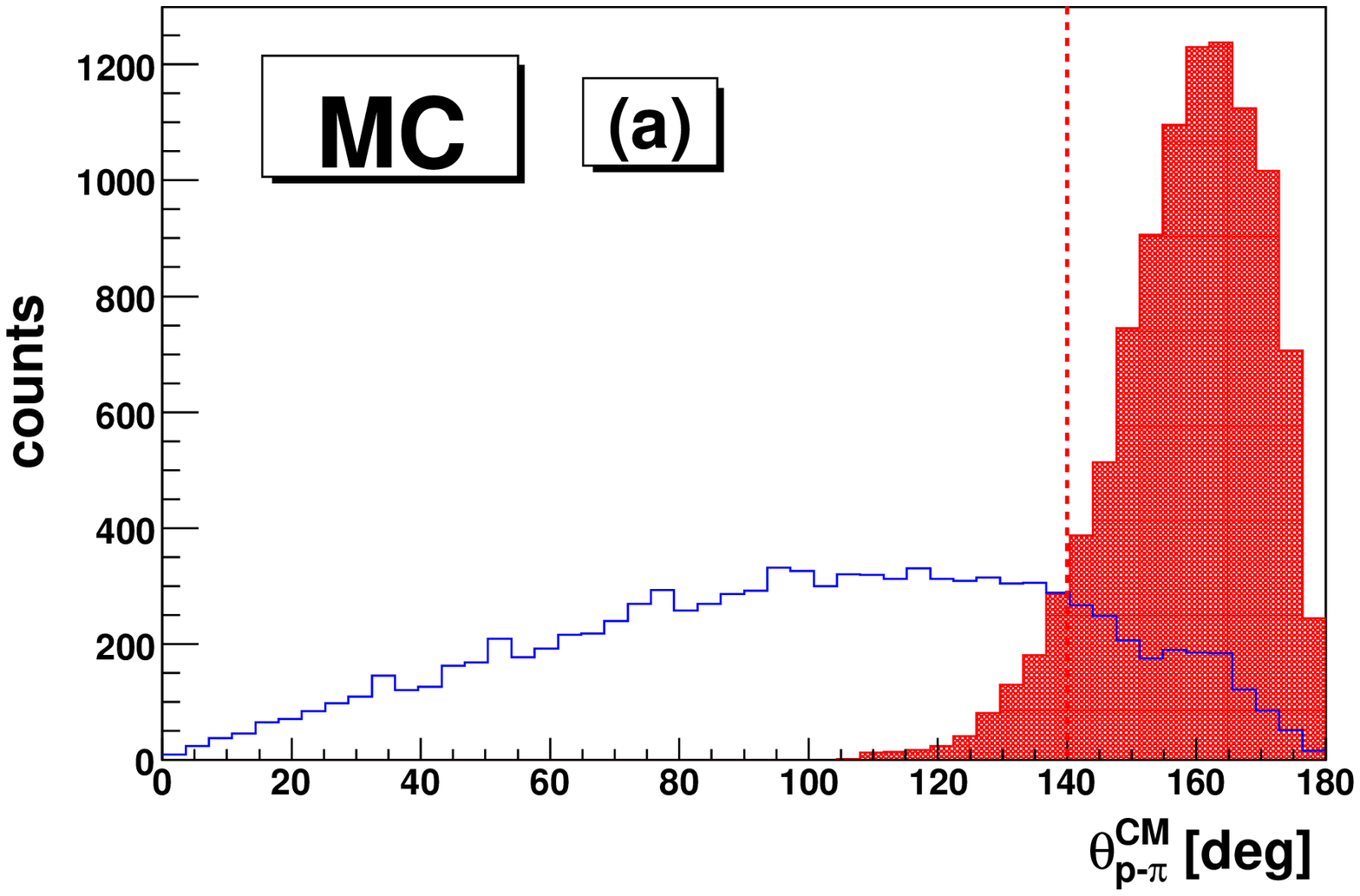}
         }
      \scalebox{\scaleFactor}
         {
              \includegraphics{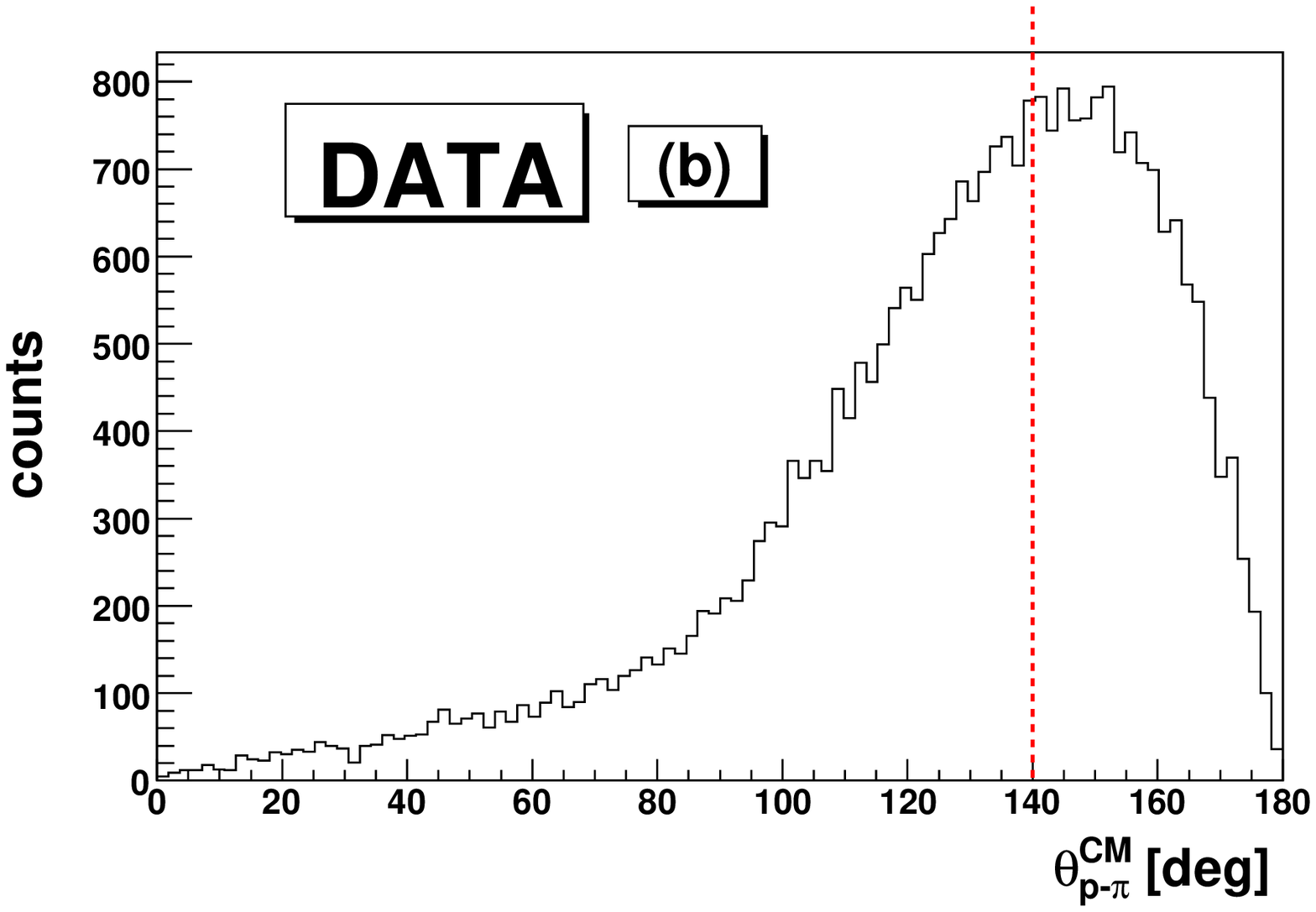}
         }

\caption{\label{proton_pion_openAnglec.m._p}(Color online) 
(Panel a)  Simulated distribution of the  $p-\pi^{-}$ opening angle in the  c.m. system for 
$dd\to(^4$He$\eta)_{bound}\to^3$He$p\pi^{-}$ reaction (red histogram) and
for the phase-space $dd\to^3$He$p\pi^{-}$ reaction (blue line).
(Panel b) Experimental distribution of the  $p-\pi^{-}$ opening angle in the c.m. system. 
In both plots the red dashed line separates the "signal-poor" and the "signal-rich" regions.
}
\end{center}
\end{figure}

In addition, further observables were taken into account in order to reduce the background.
Additional selection criteria on the $p$ and $\pi^-$ kinetic energy distributions
and the $p-\pi^-$ opening angle in the c.m. system were applied.
In the $\mbox{N}^{*}$ rest frame this angle is exactly equal to ${180^{\circ}}$
but due to the Fermi motion it is smeared by about ${30^{\circ}}$ in the reaction c.m. system (see Fig.~\ref{proton_pion_openAnglec.m._p}).
We also applied a condition on the relative $p-\pi^-$ angle in the c.m. system in the range of (140$^{\circ}$-180$^{\circ}$). 

\begin{figure}
\begin{center}
  \scalebox{\scaleFactor}
  {
    \includegraphics{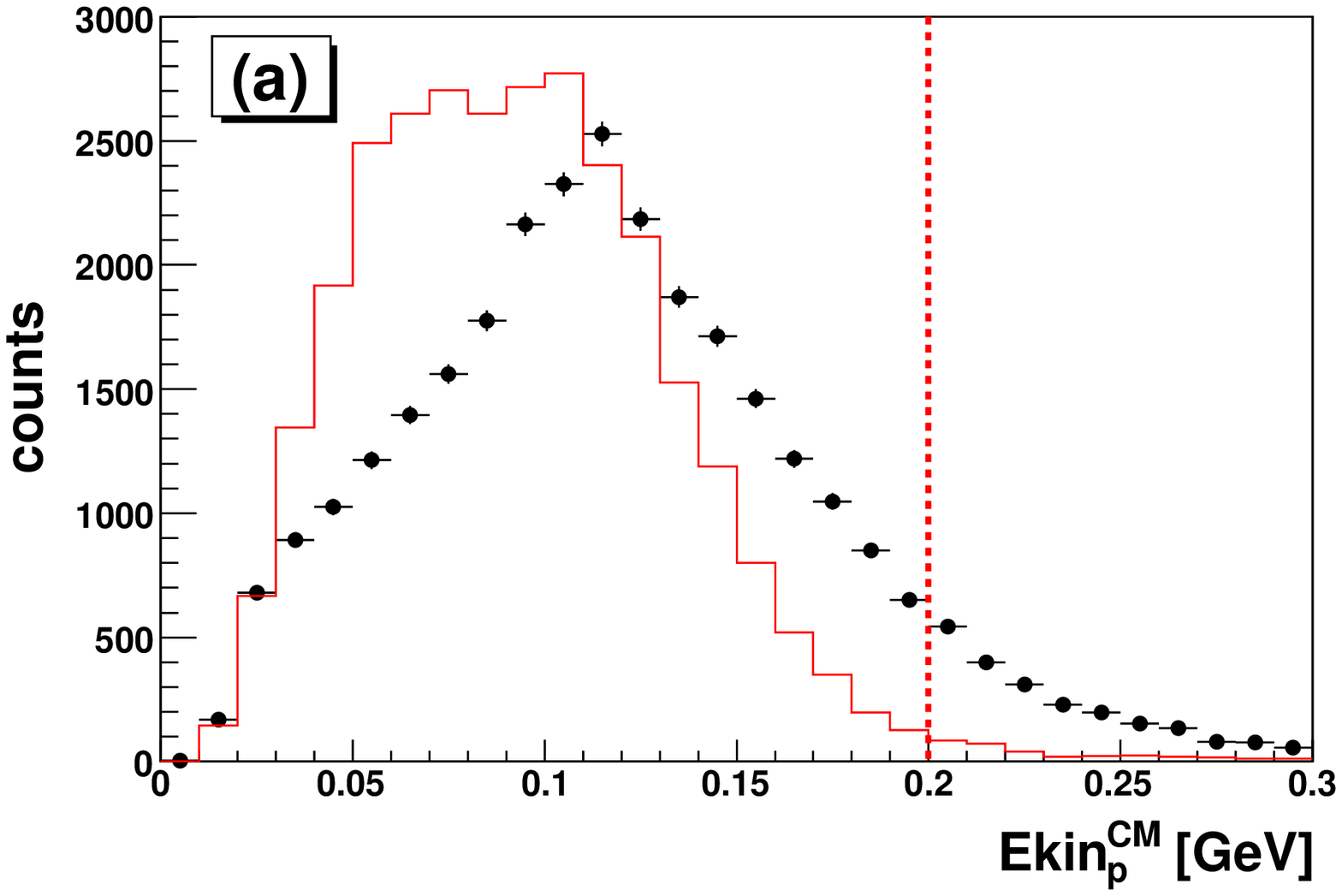}
  }
  \scalebox{\scaleFactor}
  {
    \includegraphics{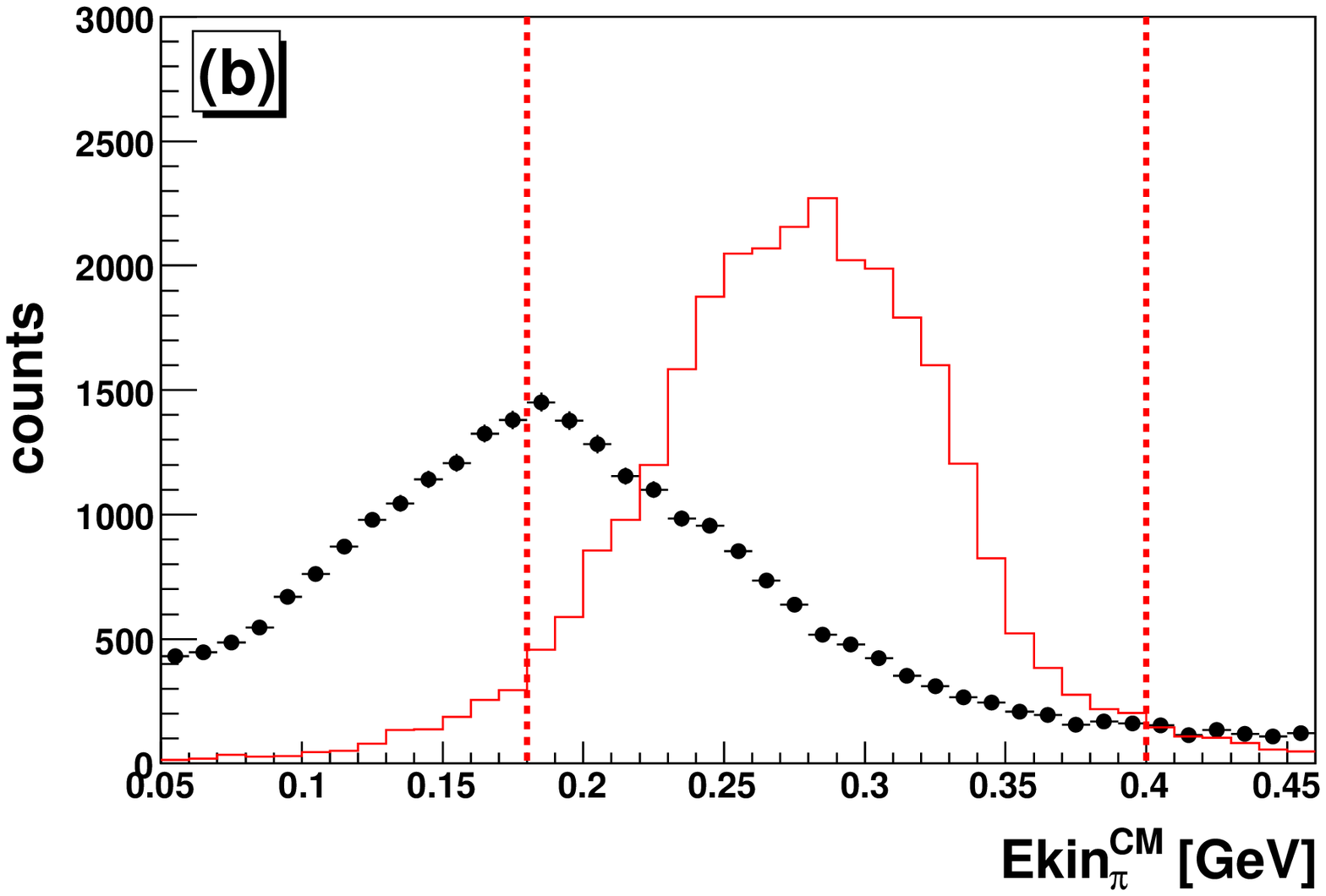}
  }
\caption[Kinetic energy distributions for $p$ and $\pi^{-}$ in the c.m. frame.]{\label{ekin_p_pim_p}(Color online) Kinetic energy distribution of protons (a) and $\pi^{-}$ (b) in the c.m. frame obtained from experiment (points) and from the simulations of a signal reaction (lines). The red dashed line indicates the boundary of the applied selection criteria: ($Ekin^{CM}_{p} < 200$ MeV, $Ekin^{CM}_{\pi^{-}} \in (180,400)$ MeV). Please note that the ranges on the X axes are different.}
\end{center}
\end{figure}

The experimental spectra of c.m. kinetic energies of protons and pions 
are compared in Fig.~\ref{ekin_p_pim_p} to the distribution expected for the signal reaction 
$dd \rightarrow (^4$He$-\eta)_{bound} \rightarrow ^3$He$ p \pi^{-}$.
For further analysis we selected the kinetic energy of protons smaller than 200 MeV  and of pions from the (180, 400) MeV interval.

After the application of the described conditions 
the number of selected events in each excess energy (Q) interval was divided
by the corresponding integrated luminosity and corrected for the reconstruction efficiency. 

The absolute value of the integrated luminosity in the experiment was determined 
using the $dd \rightarrow ^3$He$ n$ reaction and 
the relative normalization of points of the $dd \rightarrow ^3$He$ p \pi^-$ excitation
function was based on the quasi-elastic proton-proton scattering~\cite{wkPhD}.

\begin{figure}
\begin{center}
      \scalebox{\scaleFactor}
         {
              \includegraphics{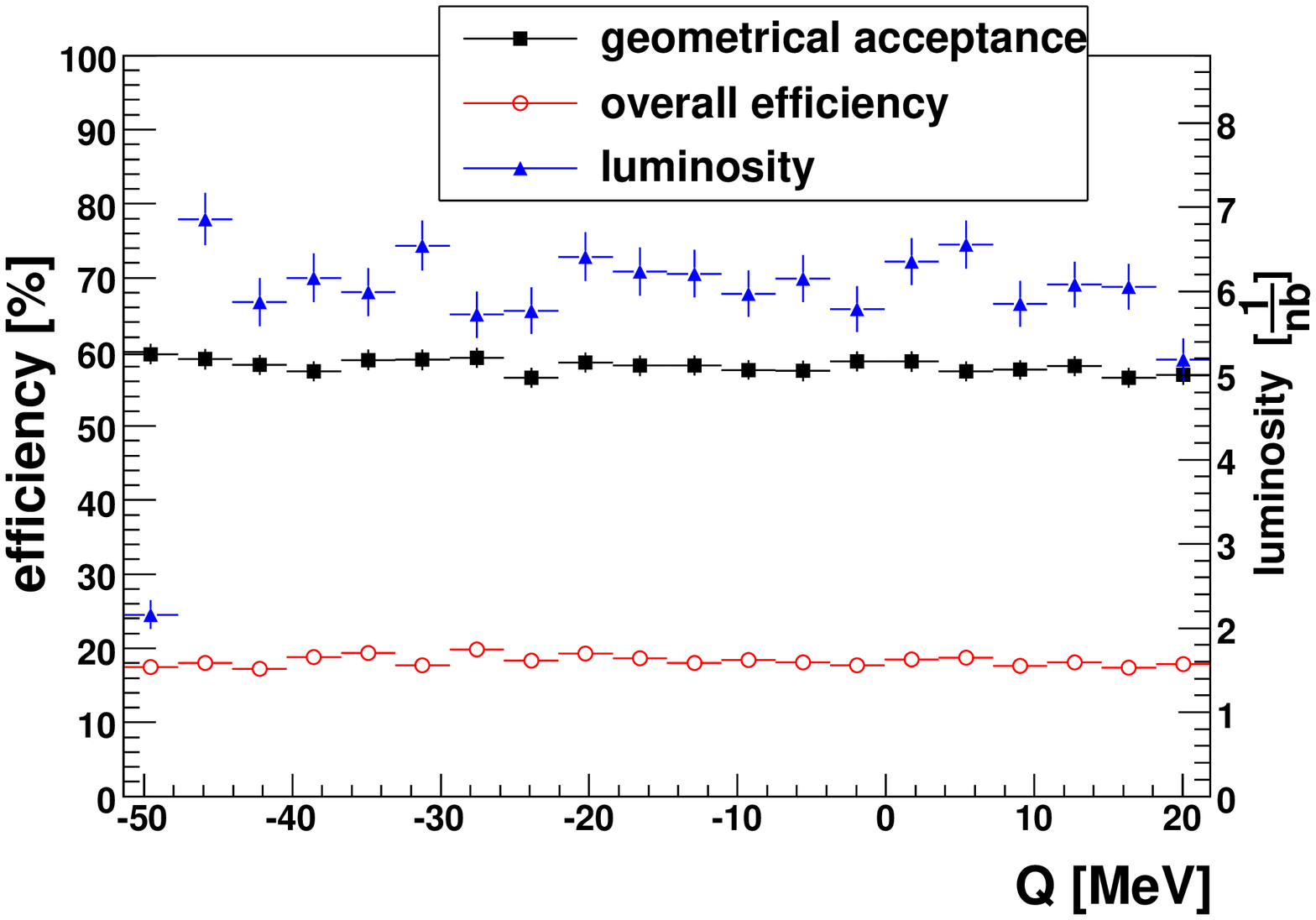}
         }
\caption{\label{efficency_and_lumin_in_Q_v2_p}(Color online) Geometrical acceptance (full black squares), overall efficiency (open red circles) and luminosity (full blue triangles) as a function of the excess energy. The right axis denotes the luminosity.}
\end{center}

\end{figure}
The luminosity as a function of the excess energy is shown as triangles in Fig.~\ref{efficency_and_lumin_in_Q_v2_p}, is flat within the statistical uncertainties. 
The geometrical acceptance is about 60\% and the overall efficiency including all selection conditions applied in the analysis is about 18\% along the whole excess energy range. 
It is important to stress that both acceptance and efficiency are smooth and constant over the studied range.

\begin{figure}
\begin{center}
      \scalebox{\scaleFactor}
      {
         \includegraphics{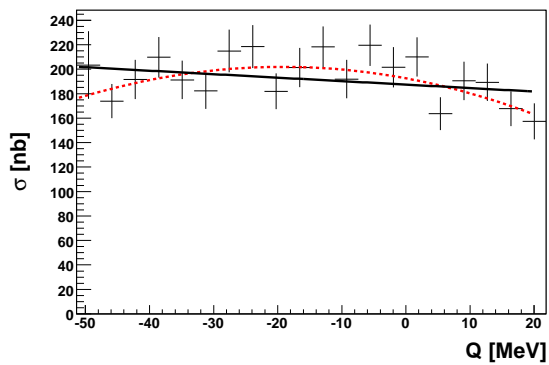}
      }
\caption[Breit-Wigner fit to the normalized excitation function]{\label{fit_10_10_allcuts_p}(Color online) 
Experimental excitation function for the $dd \rightarrow ^3$He$ p \pi^{-}$ reaction obtained after the normalization of the events
selected in individual excess energy intervals by the corresponding integrated luminosities.
The dotted  and solid lines correspond to the second  and the first order polynomials fitted to the data.}
\end{center}
\end{figure}

The excitation function obtained after the selection criteria on energy and opening angles,
the correction for the efficiency and the normalization to the luminosity  
is presented in Fig.~\ref{fit_10_10_allcuts_p}.
It can be  well described by a second order 
polynomial (dashed line) resulting in a chi-squared value per degree of freedom of 0.98 
and slightly worse by a straight line (solid line).
As in the intermediate stage of the analysis (Fig.~\ref{hExcitFuncc.m._mom_exp_bad_p}),
the final excitation function exhibits no structure which could be interpreted
as a resonance originating from the decay of the $\eta$-mesic $^4$He.

\section{Upper limit for the  $dd \rightarrow (^4{\mbox{He}} \eta)_{bound} \rightarrow ^3$He$p\pi^{-}$ cross-section}

Since no signal originating from the formation of the $^4$He$-\eta$ bound state was observed,
we estimate an upper limit for its production via the $dd \rightarrow (^4{\mbox{He}} \eta)_{bound} \rightarrow ^3$He$p\pi^{-}$ reaction.

We assumed that a signal from the bound state in the excitation curve  determined as a function of the excess energy $Q$ 
with respect to the $^4$He$-\eta$ threshold can be described by a Breit-Wigner shape:
\begin{equation} \label{bw_eq}
\sigma(Q,E_{BE},\Gamma,A)=\frac{A \cdot (\frac{\Gamma}{2})^2}{(Q-E_{BE})^2 +(\frac{\Gamma}{2})^2},
\end{equation}
where $E_{BE}$ is the binding energy,  $\Gamma$ is the width of the bound state and $A$ is the cross-section at the central energy ($Q=E_{BE}$)
for the $dd \rightarrow (^4$He$-\eta)_{bound} \rightarrow ^3$He$ p \pi^{-}$ reaction. 
In this way, we assume that there is no interference between the signal and the non-resonant background.
In order to determine an upper limit for the cross-section for formation of the $^4$He$-\eta$ bound state and its decay into the $^3$He$ p \pi^{-}$ channel  we fitted the excitation function 
with a  polynomial  describing the background (first and second order) combined with the Breit-Wigner function.
In the fit, the polynomial coefficients and the amplitude $A$ of the Breit-Wigner distribution were treated as free parameters.
The binding energy $E_{BE}$ and the width $\Gamma$ were fixed during the fit.  

The fit was performed for various values of the binding energy and the width representing different hypothesis of the bound state
properties. 
The binding energy and the width were varied in the range from 0 to -30 MeV   and from 5 to 35 MeV respectively. 
In each case, the extracted value of $A$ is consistent with zero within the statistical uncertainties, 
which confirms the  hypothesis of non-observation of the signal.

In order to  calculate  an upper limit for the $dd \to (^4\mbox{He}\eta)_{bound} \rightarrow ^3$He$ p \pi^-$
cross-section, the standard deviation of the A values ($\sigma_{A}$) obtained from the above described fit
were multiplied by the statistical factor $k$ equal to 1.28155 corresponding to the probability confidence level (CL) of 90\%.
The final results were obtained by averaging the upper limits derived from fits with a background 
described by the second and first order polynomials.

\begin{table}[!ht]\label{TabBW}
\begin{center}
\begin{tabular}{lllll}
 \hline
 \hline
 $E_{BE}$ [MeV] & $\Gamma$ [MeV] & $\sigma_{quad}$ [nb] & $\sigma_{lin}$ [nb] & $\frac{\sigma_{quad}+\sigma_{lin}}{2}$ [nb]\\
 \hline
-30 & 10  & 21.57 & 20.87 & 21\\
-30 & 20  & 23.38 & 21.77 & 23\\
-30 & 30  & 28.83 & 25.33 & 27\\
-20 & 10  & 22.49 & 18.09 & 20\\
-20 & 20  & 25.94 & 16.96 & 21\\
-20 & 30  & 33.58 & 18.03 & 26\\
-10 & 10  & 23.86 & 18.51 & 21\\
-10 & 20  & 27.78 & 16.73 & 22\\
-10 & 30  & 36.88 & 17.48 & 27\\
 \hline
 \hline
\end{tabular}
\end{center}
\caption{ 
The upper limit for the cross-section for the bound state formation and decay in the process
$dd \rightarrow (^4$He$-\eta)_{bound} \rightarrow ^3$He$ p \pi^{-}$,
determined on the 90 \% confidence level. 
The values were obtained from a fit of a Breit-Wigner function 
combined with first and second order polynomials ($\sigma_{lin}$ and $\sigma_{quad}$ respectively)
with different fixed values of binding energy $E_{BE}$ and width $\Gamma$. 
}  
\end{table}

\begin{figure}
\begin{center}
      \scalebox{\scaleFactor}
      {
         \includegraphics{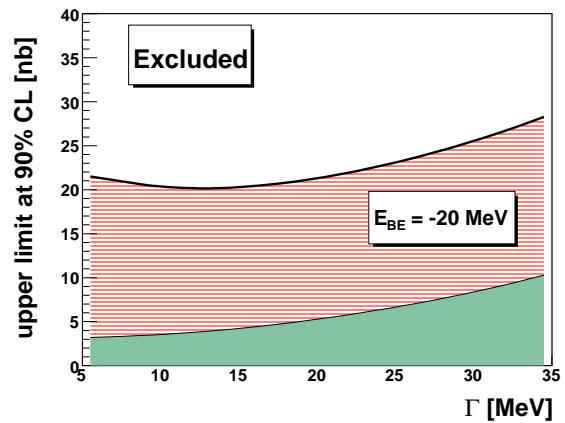}
      }
\caption{\label{upper_limit_20_BE_continous_v2_p}(Color online)
Upper limit at 90~\% confidence level of the cross-section for formation of the $^4$He$-\eta$ bound state and its decay via the $dd \rightarrow (^4$He$-\eta)_{bound} \rightarrow ^3$He$ p \pi^{-}$ reaction as a function of the width of the bound state. 
The binding energy was set to $E_{BE}$=-20\,MeV. The grey area at the bottom represents the systematic uncertainties.
}
\end{center}
\end{figure}

The examples of the obtained upper limits are given in the last column of Table~I. One can notice that these limits depend mainly on the width of the bound state but only slightly on the binding energy. The result for  $E_{BE}$=-20\,MeV is shown in Fig.~\ref{upper_limit_20_BE_continous_v2_p}.

\section{Systematics}

Systematic checks were performed by studying the sensitivity of the result to the variation 
of the selection conditions performed in the analysis  and the assumption taken in the fitting procedure.

Changing the range of the above described selection conditions within $\pm $ 10 \% gives a result consistent within the statistical uncertainties. 

The smooth reconstruction efficiency  and the luminosity dependency as a function of excess energy is of high importance because it  
eliminates the possibility of the creation of an artificial signal due to fluctuation of the acceptance or the luminosity.

Two methods were applied to extract the luminosity dependency as a function of Q. In addition to the normalization calculated on a bin by bin basis, we have estimated the luminosity dependence of Q using a fit of a first order polynomial to the data. The results of both methods are in agreement. However, an overall normalization uncertainty of luminosity is equal to 11.5 \%~\cite{wkPhD} and this value is one of the contributions to the systematic uncertainty of the upper limit. 

The description of the background shape with quadratic and linear functions 
produces additional systematic uncertainty, which was estimated as:
\begin{equation}
\delta = \frac{(\sigma_{quad} - \sigma_{lin})}{2}
\end{equation}
The systematic error grows almost linearly with the assumed bound state width from  about 5~\% ($\Gamma$ = 5  MeV, $E_{BE}$= -20 MeV) to 33~\% ($\Gamma$ = 35 MeV, $E_{BE}$= -20 MeV)  and we take that range as an  estimate of the systematic uncertainty due to the assumed shape of the background.

An important source of systematic errors comes from the Fermi momentum distribution of nucleons
inside the $^4\mbox{He}$ nucleus applied in the simulations. 
We adapted the Fermi momentum parametrization described in \cite{VH},
which is derived from the work of McCarthy, Sick and Whitney~\cite{McCarthy}. 
However, as it is shown in~\cite{Magda}, the alternative derivation of the Fermi momentum distribution done by Nogga~\cite{Nogga} is narrower by about 25 \%.
Even if the choice of the given Fermi model does not influence the experimental method, 
it affects the overall acceptance of the $^3$He ions in the Forward Detector, 
and adds an additional systematic error of 8 \%.

In principle, a complete description would require the application of a momentum distribution with 
the embedded $N^{*}$ resonance.
However, up to now, such a description of the momentum distribution is unavailable. Therefore,
we approximate this distribution by Fermi momentum distribution of nucleons inside the $^4$He nucleus.

Adding the above estimated contributions in quadrature we obtain systematic uncertainty of the upper limit which grows with bound state width from 15~\% to 36~\%. 

\section{Summary}

We performed a search for the $^4$He$-\eta$ bound state via  exclusive measurement
of the excitation function for the $dd \rightarrow ^3$He$ p \pi^-$ reaction.
The measurement was carried out with the internal deuteron beam of the COSY accelerator scattered on a deuteron pellet target 
and with the WASA-at-COSY detection system used 
for registration of the reaction products.
During the experimental run the momentum of the deuteron beam  was varied continuously within each acceleration cycle
from  2.185~GeV/c to 2.400~GeV/c, crossing the kinematic threshold for $\eta$ meson production 
in the $dd \rightarrow ^4$He$\,\eta$ reaction at 2.336~GeV/c.
This range of beam momenta corresponds to an interval of the excess energy in the $^4\mbox{He}-\eta$
system from -51.4~MeV to 22~MeV.

For the first time in the experimental search for mesic nuclei all ejectiles were measured and the reaction was identified exclusively.

No signal from $\eta$-mesic  $^4$He was observed.  
The upper limit for the cross-section for the bound state formation and decay in the process
$dd \rightarrow (^4$He$-\eta)_{bound} \rightarrow ^3$He$ p \pi^{-}$,
was determined on the 90 \% confidence level and it varies from 20~nb to 27~nb for the bound state width ranging 
from 5~MeV to 35~MeV, respectively.
The upper limits depend mainly on the width of the bound state and only slightly on the binding energy.

\section{Support}
This work has been supported by FFE funds of Forschungszetrum J\"ulich, by the European Commission under the 7th Framework Programme through the 'Research Infrastructures' action of the 'Capacities' Programme. Call: FP7-INFRASTRUCTURES-2008-1, Grant Agreement N. 227431 and by the Polish National Science Center under grants No. 0320/B/H03/2011/40 and 2011/01/B/ST2/00431, 2011/03/B/ST2/01847, 0312/B/H03/2011/40 and by the Foundation for Polish Science (MPD).

\end{document}